\documentclass[journal]{IEEEtran}
\usepackage{color,amsfonts,colortbl,amsmath,amssymb,booktabs,multirow,psfrag,bm,amsbsy,cite,float}
\usepackage{enumerate}
\usepackage{caption}
\usepackage{algpseudocode}
\usepackage{algorithmicx}
\ifCLASSINFOpdf

\usepackage{graphicx}
\usepackage{subcaption}
\usepackage{flushend}
\usepackage{dblfloatfix}
\usepackage[normalem]{ulem}

\usepackage{tikz}
\usetikzlibrary{arrows,shapes,positioning}
\usetikzlibrary{decorations.markings}
\tikzstyle arrowstyle=[scale=1]
\tikzstyle directed=[postaction={decorate,decoration={markings,
    mark=at position .65 with {\arrow[arrowstyle]{stealth}}}}]
\tikzstyle reverse directed=[postaction={decorate,decoration={markings,
    mark=at position .65 with {\arrowreversed[arrowstyle]{stealth};}}}]
    \tikzstyle normal=[postaction={decorate,decoration={markings,
    mark=at position 1 with {\arrow[arrowstyle]{stealth}}}}]

\DeclareMathOperator*{\argmin}{arg\,min}

\DeclareGraphicsExtensions{.pdf,.jpeg,.jpg,.png,.eps}\else  
\fi

\begin{document}
 

\title{The COST IRACON Geometry-based Stochastic Channel Model for
Vehicle-to-Vehicle Communication in Intersections}

\author{\IEEEauthorblockN{Carl Gustafson\IEEEauthorrefmark{1},
		Kim Mahler\IEEEauthorrefmark{2},
		David Bolin\IEEEauthorrefmark{3},
		Fredrik Tufvesson\IEEEauthorrefmark{4},
		}\\
	
	\IEEEauthorblockA{\IEEEauthorrefmark{1}SAAB Dynamics, Link\"oping, Sweden}\\	
		\IEEEauthorblockA{\IEEEauthorrefmark{2}NYU WIRELESS, New York University, Brooklyn, NY 11201, USA}\\
		\IEEEauthorblockA{\IEEEauthorrefmark{3}CEMSE Division, King Abdullah University of Science and Technology, Saudi Arabia}\\
		\IEEEauthorblockA{\IEEEauthorrefmark{4}Lund University, Dept. of Electrical and Information Technology, Lund}, Sweden\\

		email: carl.gustafson@saabgroup.com, web: https://github.com/COSTIRACONV2VGSCM}

\maketitle

\begin{abstract}
Vehicle-to-vehicle (V2V) wireless communications can improve traffic safety at road intersections and enable congestion avoidance. However, detailed knowledge about the wireless propagation channel is needed for the development and realistic assessment of V2V communication systems. We present a novel geometry-based stochastic MIMO channel model with support for frequencies in the band of 5.2-6.2 GHz. The model is based on extensive high-resolution measurements at different road intersections in the city of Berlin, Germany. We extend existing models, by including the effects of various obstructions, higher order interactions, and by introducing an angular gain function for the scatterers. Scatterer locations have been identified and mapped to measured multi-path trajectories using a measurement-based ray tracing method and a subsequent RANSAC algorithm. The developed model is parameterized, and using the measured propagation paths that have been mapped to scatterer locations, model parameters are estimated. The time variant power fading of individual multi-path components is found  to be best modeled by a Gamma process with an exponential autocorrelation. The path coherence distance is estimated to be in the range of 0-2 m. The model is also validated against measurement data, showing that the developed model accurately captures the behavior of the measured channel gain, Doppler spread, and delay spread. This is also the case for intersections that have not been used when estimating model parameters.    
\end{abstract}

\begin{IEEEkeywords}
GSCM, V2V, V2X, Channel model.
\end{IEEEkeywords}

%
\IEEEpeerreviewmaketitle


\section{Introduction}
\IEEEPARstart{V}{ehicle}-to-vehicle (V2V) communications has potential to improve road safety through collision avoidance systems and can help to enable an improved traffic flow with congestion avoidance. Vehicles such as trucks and cars are nowadays equipped with numerous sensors, and can share important information between each other if they are connected by wireless links. The research interest in this area  was originally sparked by the 75 MHz band allocated at 5.9 GHz by the regulator FCC in the US and by the IEEE 802.11p standard \cite{802}. More recently, research has been conducted exploring the possibilities of using LTE or 5G technologies for communication between vehicles \cite{5G}. The LTE-V standard is nowadays an alternative to the 802.11p standard. It includes mode 3, which support V2V communication aided by cellular resource allocation, and mode 4 which does not require any cellular connection \cite{V2X}. Systems involving base-stations might be limited by latency, which is critical in safety systems, and might also be limited by poor coverage and blind spots in certain areas. Vehicles are thus expected to be equipped with dedicated transceivers so that communication is enabled even in spots with poor base station coverage. 

Future V2V wireless applications are numerous, and several of them will need to rely on a secure and reliable channel with low latency. Some future applications might also need high data rates. For these reasons, multiple-input, multiple-output (MIMO) technologies are likely to be employed. MIMO techniques can support higher data rates through spatial multiplexing, it can improve resistance to fading through diversity and also opens up the door to accurate radio-based localization and positioning techniques. In order to develop next generation wireless systems for vehicles, detailed information about the propagation channel is needed. Although a lot of work has already been done in this area, no work has truly captured the multi-path channel behavior in urban environments. In this paper, we therefore focus our attention to intersections in urban environments. These types of environments are important from a safety point of view, since the visual line of sight often is blocked, and many accidents occur there \cite{Crash}. Radar- or camera-based collision avoidance systems might also have a poor performance, as they have limited capabilities of "seeing" around the corner. 

Several papers have already characterized the properties of wireless channels in urban intersections, and have presented the general behavior of packet error rates, channel gains, eigenvalue distributions and delay and Doppler spreads \cite{char1,Char2,char3}. Mangel et. al. have developed a path loss and fading model, based on measurements in representative intersections in the city of Munich, Germany \cite{Mangel}. Abbas et. al. extended this model by adding an intersection dependent parameter, and then validated this model against real-world intersection measurements in the cities of Malm\"o and Lund, Sweden \cite{Abbas}. More recently, Nilsson et. al. presented a path loss and fading model for the multi-link case in urban intersections, based on measurements in the city of Gothenburg, Sweden \cite{Mikael}. The models in \cite{Mangel,Abbas,Mikael} are simple and easy to use, but provides no means of simulating the spatio-temporal MIMO channel. Also, it is unclear how well these models perform in more general urban environments.

For highway and rural scenarios, a geometry-based stochastic channel model (GSCM) has been developed by Karedal et al. \cite{Karedal}. GSCMs can capture the non-stationary spatio-temporal behvaior of dynamic wireless channels both accurately and efficiently, making it an ideal candidate for V2V channels. It does so by combining simplified ray tracing methods with a stochastic description of scatterer locations and properties. This enables a fast simulation of non-stationary MIMO channels, and also supports simulation of arbitrary antenna patterns and array configurations. However, the highway model in \cite{Karedal} does not include propagation effects that are vital for urban scenarios, such as obstruction and diffraction and higher order interactions. 
A few GSCMs for urban V2V scenarios have been presented in the literature \cite{gscm1,gscm2, Boban}. In\cite{gscm2} the parameter estimates for highway scenarios in \cite{Karedal} are applied, and the effects of building obstructions are added. The theoretical model in \cite{gscm1} is based on multi-path clusters placed along walls and building corners.  A V2V channel model for large scale simulations (including urban scenarios) is presented in \cite{Boban}. It is a geometry-based model which includes reflection, diffraction and paths that are obstructed by buildings or foliage. Large scale signals are calculated deterministically, whereas the small scale fading of the received power is determined stochastically. The models in \cite{gscm1,gscm2, Boban} are only validated against data of large scale parameters such as received power. 

To the author's best knowledge, we present the first V2V channel model for urban scenarios based on measured and highly resolved multi-path components. The aim of this paper is to be able to accurately model multi-path behavior in challenging V2V scenarios, in order to enable improved V2V MIMO techniques and V2V positioning and localization techniques. We have developed a non-stationary geometry-based stochastic MIMO channel model for arbitrary urban environments, based on extensive  measurements in urban scenarios. The model supports a frequency range of 5.2-6.2 GHz, and the modelled multi-path channel behavior is validated against measurement results. We extend existing GSCMs \cite{Karedal,gscm2}, by including the effects of i) higher order interactions, ii) obstructions by buildings, foliage and other objects, iii) diffraction around corners, and by iv) prescribing scatterers with a non-isotropic angular gain function. 

Our generic model supports simulations of arbitrary vehicular environments. The model is parameterized based on high resolution measurements performed in four different real-world urban intersections in the city of Berlin, Germany. Model parameters are estimated from two different intersections: a narrow and an open intersection. The model performance is then validated by comparing simulated channels with the measured ones. This is done for the narrow and open intersection, as well as for a wide and a T-shaped intersection. The spatio-temporal behavior of the simulated channels agree well with the measured channels, and the peak PDP power, mean delay and RMS delay spread can be predicted quite well by the model. The simulated Doppler-delay profiles also agree well with the measured ones. 
The generic model is also applicable in other areas. While the presented model is aimed at V2V simulations at 5.9 GHz, the generic model framework could also be beneficial when modelling dynamic propagation channels above 6 GHz. Channels at mm-wave frequencies are of special interest, as they are heavily influenced by obstructing objects, and need to rely on directional beamforming. Hence, the directional aspect of the scattering objects need to be included. 
 
\section{Measurement Campaign}
\subsection{Measurement Equipment}

The radio channel data was collected using the HHI channel sounder, a wideband measurement device developed at the Fraunhofer Heinrich Hertz Institute (HHI) with a bandwidth of 1 GHz at a carrier frequency of 5.7 GHz \cite{X1}. The measurement bandwidth enables highly resolved MPC trajectories, due to the delay resolution of 1 ns. The channel sounder consists of a transmitter unit and a receiver unit, which are installed in conventional passenger vehicles. The measurement vehicles are equipped with two vertically polarized omnidirectional antennas that are mounted on the roof at the left and right edges of the vehicle. 
To record the position of the vehicles during measurements, we used the positioning system GeneSys ADMA-G, which works reliably and is highly accurate even in deep street canyons with limited GPS satellite coverage. The measurements were accompanied by conventional video cameras mounted on the inner windshield of both vehicles and video cameras mounted on the roof of both vehicles. A continuous voice link between the drivers ensured a collision course of the vehicles during measurement. 

\subsection{Measurement Scenario}
The measurement campaign included a vast number of different intersections in the city of Berlin. The measurements were taken on a weekday during the day, roughly between 10:00 and 16:00. This represents a scenario with moderate traffic, as it was not measured at rush-hour nor at night or weekends. In this paper, we limit our work to the measurements conducted in the four different intersections which are detailed in Table I. These intersections were selected because they represent a wide range of intersection types. Specifically,  narrow and wide four-way intersections with buildings in all four corners, an open-type intersection, and a T-shaped intersection. 

All intersections are characterized by a massive obstruction between the communicating vehicles. Both vehicles drive in street canyons, which allow scattered waves to travel from transmitter to receiver.  According to \cite{Mangel}, intersections with buildings in four corners account for 70-90\% of all 4 leg intersections  and are grouped as Category 2 in \cite{Char2}. For this paper, we have analyzed three different measurement runs per intersection. The distance to the crossing center was up to 200 m and the speed of the vehicles approaching the intersection varied depending on traffic circumstances between 30-60 km/h. Aerial photos of the selected intersections are shown in Fig.~1.

\begin{figure*}[t]
        \centering
        \includegraphics[width=0.24\textwidth]{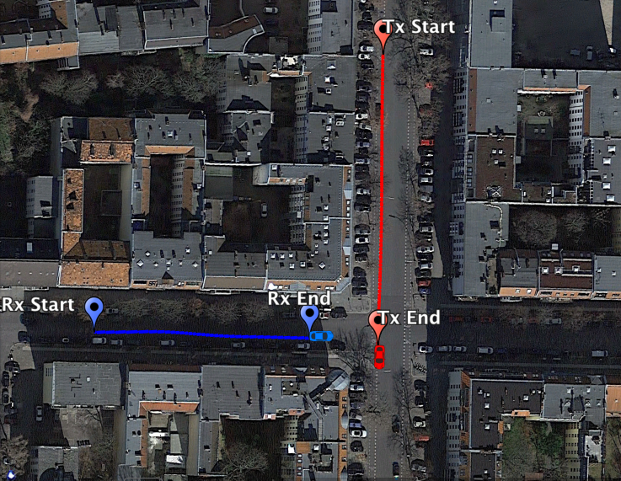}%
        \hfill
        \includegraphics[width=0.24\textwidth]{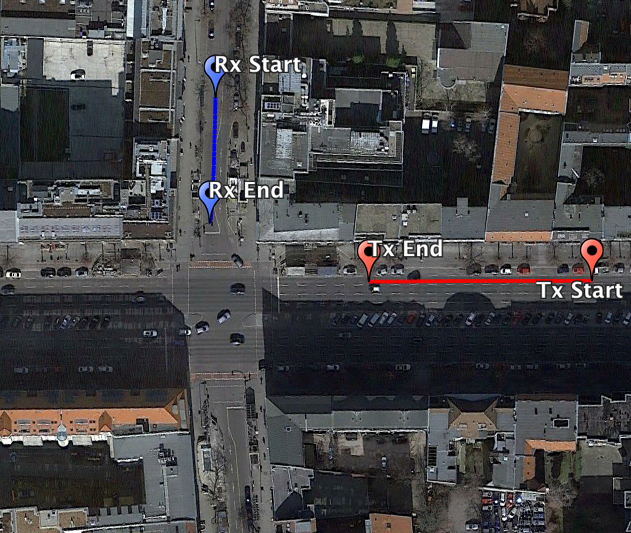}
        \hfill
        \includegraphics[width=0.24\textwidth]{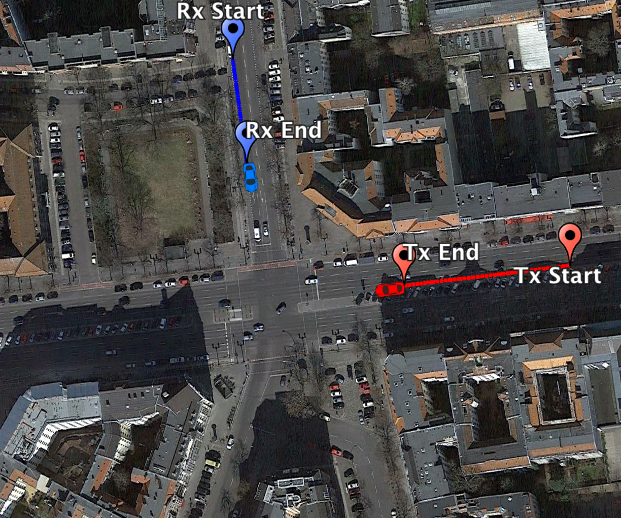}
                \hfill
        \includegraphics[width=0.24\textwidth]{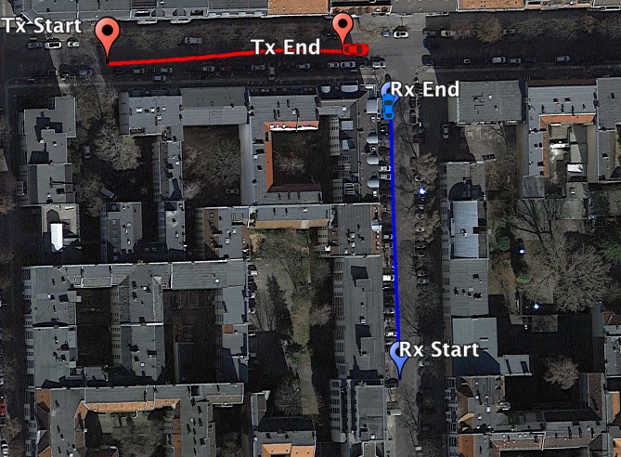}
         \caption{Aerial photos of the four different intersections where measurements took place. Form left to right: the narrow, wide, open and T-shaped intersections. Trajectories of the Tx and Rx car for a single measurement run are also shown in each image.}
         \label{fig:aerial}
\end{figure*}

\begin{table}[h]
\caption{Overview of investigated intersections}
\begin{center}
\begin{tabular}{|c|c|c|c|} \hline
Street &  & Crossing center  & Intersection  \\ 
names & Runs & position (lat/lon) & category \\ \hline
Pestalozzistr. &  & 52$^{\circ}$30.456'  & \\
- Schl\"uterstr. & 3  & 13$^{\circ}$19.064' & Narrow \\ \hline 
Wilmersdorferstr. &  &52$^{\circ}$30.697'   & \\
- Bismarckstr. & 3  & 13$^{\circ}$18.310'  & Wide \\ \hline 
Pestalozzistr. &  &52$^{\circ}$30.462'   & \\
- Wielandstr. & 3  & 13$^{\circ}$18.964'  & T-shaped \\ \hline 
Schlossstr. &  &52$^{\circ}$30.664'   & \\
- Bismarckstr. & 3  & 13$^{\circ}$17.838'  & Open \\ \hline 
\end{tabular}
\end{center}
\label{default}
\end{table}%


%
%

\section{Post-processing}
In order to develop a geometry-based stochastic channel model, it is necessary to identify how each multi-path component (MPC) evolves as a function of time with respect to delay and power. The time-delay characteristics of each MPC can then be associated with different scattering objects in the environment. 
 
\subsection{Initial scatterer identification}
  The so-called measurement-based ray tracing method \cite{Poutanen} is effective at mapping measured multi-path trajectories to geometrical objects. This method aims at reconstructing a channel measurement run in a computer simulation, and is done by comparing ray tracing results with measured data. In order to reflect the real-world conditions of the propagation process, the ray tracing simulation has to include all relevant geometrical information of the measurement environment. The intersection geometry model has been automatically derived from a data set provided by the city of Berlin, which includes the exact position of house walls, sidewalks and trees. In order to complete this geometrical model, all remaining relevant objects, such as traffic signs or street lamps, have been measured accurately with a laser distance meter on-site and included in the geometry model manually. The GNSS trajectories of the measurement vehicles are transformed from the geographic coordinate system (latitude, longitude) to the Cartesian coordinate system ($x,y$), where the origin of the Cartesian system is the intersection center. In order to accurately simulate the MIMO propagation channel, the positions of the simulated antennas are shifted accordingly.

The association of MPC tracks and scatterer locations is based on an evaluation of delay, Doppler, angle estimates, and the delay characteristic. To ensure a robust association, a so-called semi-automated reasoning method is employed, which involves a limited number of automated suggestions and a subsequent human decision. This means that for each measured MPC, the automated algorithm presents the human editor with the closest ray tracing candidates in terms of the delay and Doppler. The geometrical positions of these scatterer candidates are then assessed against the estimated path angles. The lifetime of the measured MPC is compared with possible obstruction effects, due to the geometrical circumstances of the scattering candidates. Finally, the measured MPC and the simulated ray tracing candidates are compared in terms of their change of Doppler over time, which is directly related to the change of the path angles over time. More details on this can be found in \cite{MahlerTracking}.  

\subsection{Identifying individual sub-components}
The measurement-based ray-tracing method gives very robust results, and can accurately associate the overall scatterer locations with the measured time-delay tracks. For each path, $s$, the data is indexed by $i=1,2,\ldots$, describing the measured path propagation distance, $d_i$, at time instants $t_i$. The cartesian coordinates for the initial scatterer locations for path $s$ is given by $x_{s,o}$ and $y_{s,o}$, where $o$ is the index for the order in which the interactions take place. Many of the initially identified paths consist of several separable paths. So, for each initial scatterer, the results are refined by identifying sub-paths within each initial path. This is done by a separate algorithm, as the ray-tracing based method cannot easily distinguish between multiple scattering objects that are very close to each other. Due to the paths being close to each other, and not being visible across all antenna element combinations at the same time, it is not possible to utilize the information from the angular and Doppler domain when identifying these sub-paths. Instead, we rely on refining the scatterer locations based on the path propagation distance over time, and by constraining the possible scatterer locations to be in close vicinity of the scattering object identified originally. To find $J$ subpaths of interaction order 1, with scatter locations $s_j = (x_j, y_j), j=1\ldots, J$, we want to solve the minimization problem 
\begin{eqnarray}\label{eq:loss}
\argmin_{s,z}\left[\sum_{i=1}^n\sum_{j=1}^J 1(z_i = j) \left(d_i-d_j(t_i)\right)^2\right],
\end{eqnarray}
where $d_i$ is the measured propagation distance for observation $i$, and $d_{j}(t_i)$ is the modelled propagation distance for the $j$th subpath at time $t_i$. Furthermore, $s=\{s_j\}_{j=1}^J$ is the set of unknown scatterer locations, and $z=\{z_i\}_{i=1}^n$ is a set where $z_{i}$ is a variable determining which subpath the $i$th observation belongs to. Lastly, $1(z=k)$ denotes an indicator function with $1(z=k)=1$ if $z=k$ and $1(z=k)=0$ otherwise. The modelled propagation distance for a first order interaction is:
\begin{align*}
d_{j}(t)^2 &= \|s_{Tx}(t)-s_j\|^2 +  \|s_j - s_{Rx}(t)\|^2.
\end{align*}
To find subpaths with a higher interaction order $o$, we solve a similar minimisation problem as \eqref{eq:loss}, but where we have to find $o$ different scatter locations, $s_{j1},\ldots, s_{jo}$, for each subpath, and the distance of a specific configuration of scatter locations is modified accordingly. For example with $o=2$ we have 
\begin{align*}
d_{j}(t)^2 &= \|s_{Tx}(t)-s_{j1}\|^2 +  \|s_{j1} - s_{j2}\|^2 + \|s_{j2} - s_{Rx}(t)\|^2.
\end{align*}
Solving the minimziation problem \eqref{eq:loss} could be thought of as estimating a mixture non-linear regression model to the data, which for example could be done using the EM algorithm. However, any such likelihood-based method will be highly sensitive to starting values and we therefore instead use the following Random sample consensus (RANSAC) algorithm for the subpath estimation. We start by deciding the number of subpaths $J$ through visual inspection (this was deemed to be sufficient since the identified paths are clearly separated). Then we estimate $z$ and $s_1,\ldots, s_j$ iteratively as follows. 
\begin{enumerate}
\item Define $Y$ as the set of all $n$ observation pairs $Y_i = (t_i, d_i)$, set $j=1$, and $n_j = 0$.
\item Randomly select ten distinct indices $i_1,\ldots, i_{10}$ in $\{1,\ldots, n\}$, estimate a scatter location 
\begin{eqnarray}
\tilde{s} = \argmin_{s}\left[\sum_{\ell=1}^{10} \left(d_{i_\ell}-d_j(t_{i_\ell})\right)^2\right],
\end{eqnarray}
and define the corresponding distance function $\tilde{d}(t)$. 
\item Calculate the number of observation pairs in $Y$ with a distance at most $0.3$ away from $\tilde{d}(t)$:  
$$
\tilde{n} = \sum_{j=1}^n 1(|\tilde{d}(t_i) - d_j|< 0.3).
$$
\item If $\tilde{n} > 0.05n$ and $\tilde{n} > n_j$, set $n_j = \tilde{n}$ and $s_j = \tilde{s}$. 
\item Repeat steps 2-4 $500$ times.
\item Set $d_j(t)$ as the distance function corresponding to the scatter location $s_j$ and let $I_j$ denote the set of indices for all observation pairs which have a distance smaller than $0.45$ to $d_j(t)$. Re-estimate $s_j$ based on this data: 
\begin{eqnarray}
s_j = \argmin_{s}\left[\sum_{\ell\in I_j}\left(d_{i_\ell}-d_j(t_{i_\ell})\right)^2\right].
\end{eqnarray}
\item Decimate the data set by excluding the points that were used in step 6: Define $Y = \{Y_i\}_{i \notin I_j}$, set $n$ to the number of observation pairs in $Y$. 
\item If $j < J$ and $n\geq10$, increase $j$ by one, set $n_j = 0$, and go to step 2. Otherwise stop the estimation.
\end{enumerate}
In our case, after the final estimation, there are remaining data points, but they are all very weak, and are likely attributed to diffuse scattering interactions. This data is discarded, and is instead modelled as diffuse interactions. For a more general case, the above RANSAC method needs to be extended in order to take care of cases when there are remaining data points of significance, or when all data points have been decimated before reaching $R$ sub-paths. This can for example be done by changing the thresholds $0.3$ m and $0.45$ m in the algorithm. The choice of using 0.3 m in our case is motivated by the fact that the delay resolution is about 0.3 m (slightly better as the measurement data has been oversampled). Empirically, we found that the best fit was found when using 0.3 and 0.45 m, respectively. One could also use the result from the algorithm as a starting value for an EM algorithm to estimate a full mixture regression model to the data. This was however not deemed necessary in our case.  

\begin{figure}[h]
	\centering
	\includegraphics[width=0.47\textwidth]{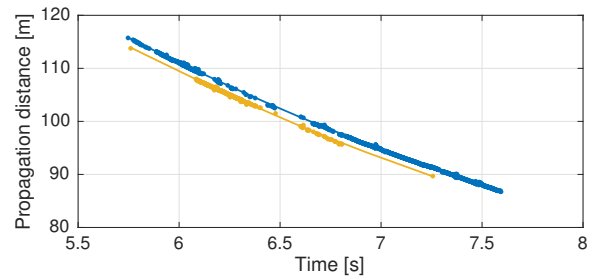}
	\caption{Typical result from the RANSAC algorithm. Two distinct sub-components have been identified, as indicated by the blue and the yellow dots. The blue and yellow lines represent the modelled propagation distance of the estimated locations of the scatterers.}
	\label{fig:RANSAC}
\end{figure}

Fig.~\ref{fig:RANSAC} shows results from the RANSAC algorithm, where two paths have been identified. The RANSAC algorithm fine-tunes the results from the initial scatterer association. The updated scatterer locations are in close vicinity of the previously identified scattering objects, and makes sense from a propagation point of view. Scatterer locations are fixed, unless the scatterer is a moving object, such as a car. According to the measurement data, there are very few components that are interacting with moving objects, and when it happens, it is usually a result of large vehicles such as vans or trucks.
\setcounter{figure}{3}
\begin{figure*}[b!]
    \centering
        \includegraphics[trim=25 0 45 5,clip,width=0.24\textwidth]{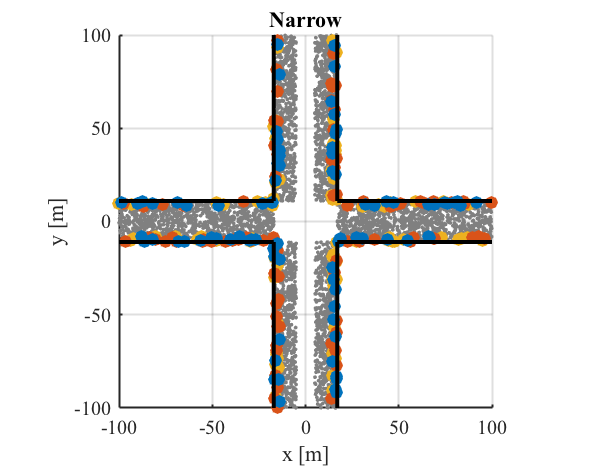}
        \hfill
        \includegraphics[trim=25 0 45 5,clip,width=0.24\textwidth]{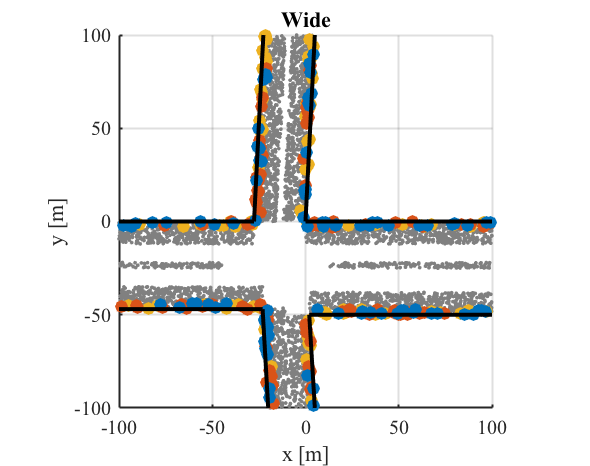}
        \hfill
        \includegraphics[trim=17 0 53 5,clip,width=0.24\textwidth]{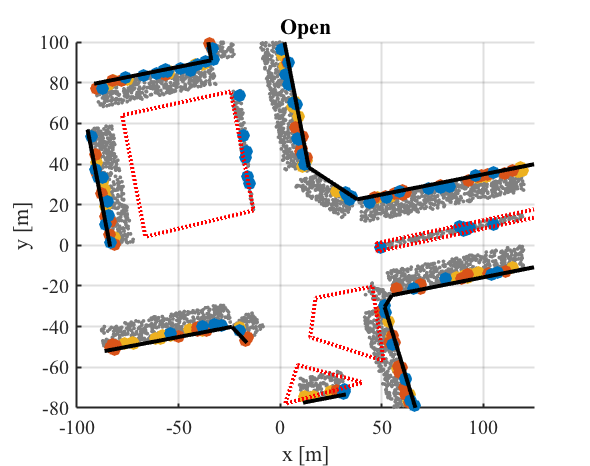}
        \hfill
        \includegraphics[trim=18 0 52 5,clip,width=0.24\textwidth]{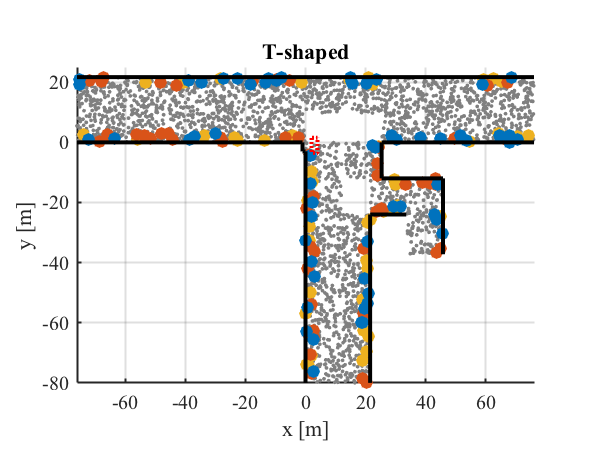}
    \label{fig:maps}
      \caption{GSCM models of the intersections, with a random realization of the scatterer locations. The blue, red, yellow and gray dots represent first-order, second-order, third-order and diffiuse scatterers, respectively. Areas enclosed by dashed red lines are affected by extra attenuation due to foliage and other objects.}
\end{figure*}

\section{Geometry-based Channel Model}
Based on the association of multi-path tracks from the measurement to point scatterers in the environment, we are now able to derive and parameterize the GSCM model. In this section, we first present experimental results of the distribution of point scatterers in the environment and show how this can be modelled. Then we present a novel model for the spatially dependent multi-path power, including the effects of building obstructions, diffraction around corners and penetration losses for areas with trees, foliage and other objects. Lastly, the power fading is found to be appropriately modelled by a Gamma process.

\subsection{Scatterer distribution}
The locations of scattering interactions along building facades that have been identified by the RANSAC algorithm is shown in Fig.~\ref{fig:scattmap}. Additional scattering interactions that occur farther away from the building facades have also been identified, but they have been omitted in the figure for the sake of clarity, and owing to their relative weak power. 
\setcounter{figure}{2}
\begin{figure}[h]
	\centering
	\includegraphics[width=0.45\textwidth]{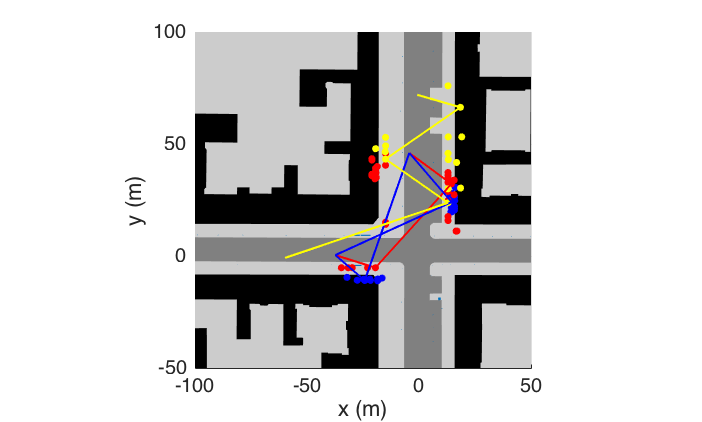}
	\caption{Identified locations for scattering interactions along building facades. Depicted are first-, second and third order interactions, shown as blue, red and yellow dots, respectively. In the figure, black areas indicate buildings, light grey areas indicate pavement/courtyards and dark grey areas are roads.}
	\label{fig:scattmap}
\end{figure}

\setcounter{figure}{4}
It can be noted that some scatterer locations appear to be located behind the facade. This might be attributed to in-building interactions, inaccuracy of the map data, inaccuracy in the GNSS data of the car positions and/or interactions that occur with a non-zero elevation angle. Fig.~\ref{fig:scattmap}, also shows some lines indicating the assumed propagation path for some of the identified first, second and third order interactions. This illustrates that scatterers seem to appear when they are \emph{unobstructed} and have \emph{similar incoming and outgoing angles} with respect to the wall surface normal. This behavior is captured by the model and is described {by Eq. (\ref{eq:ga}). We also note that in some other measurement campaigns, single bounce interactions stemming from corners such as the south-eastern one in Fig.~\ref{fig:scattmap} have been observed. In the data from the measurement campaign used in this paper, few such interactions have been identified. This could be a result of the sharp corner usually encountered in this specfic campaign, and it is likely that buildings with bevelled or rounded corners show a more significant contribution.  
From the identified scatterer locations, it is possible to estimate the geometrical distribution of the scatterers. When doing so, only the areas that are visible during the measurement run are considered. These empirical results are not shown for the sake of brevity, but indicate that scatterers are distributed approximately uniformly along the visible parts of the wall surface, and uniformly with a certain width out from the wall surface. The width of the bands are estimated to be in the range from 2.4 to 3.0 m. For each visible area, the intensity of the number of scatterers per m$^2$, is found to be $\chi_{1}=0.052$ m$^{-2}$, $\chi_{2}=0.045$ m$^{-2}$, and $\chi_{3}=0.03$ m$^{-2}$, for first-, second-, and third-order interactions, respectively.

We now assume that the scatter locations can be modelled as occurring uniformly in bands along each \emph{entire} wall, not just in certain visible areas. Scatterers are instead rendered visible or not solely based on building obstructions and the spatial gain assigned to each scatterer. This is described in detail in Sec. \ref{sec:pathgain}. 
In our GSCM model, first-, second- and third-order scatterers are placed on the map in certain areas based on the estimated scatterer locations. Scatterers tend to appear along building walls, pavements and areas with parked cars. Fig.~4 shows the GSCM model for the measured intersections, with a random realization of the scatterer locations. In our model, the scatterers are placed on the map in the following way: 
 \begin{enumerate}
 \item Draw first, second and third order scatterers uniformly over the entire map according to the respective intensities, $\chi_{1}$, $\chi_{2}$, and $\chi_{3}$. 
 \item For each wall segment in the scenario, define a scattering area with four corners given by $\mathbf{p}_{0}$, $\mathbf{p}_{1}$, $\mathbf{p}_{0}+\hat{\mathbf{n}}W$, $\mathbf{p}_{1}+\hat{\mathbf{n}}W$, where $W$ is the width of the scattering area, $\mathbf{\hat{n}}$ is a unit normal vector pointing out from the wall surface and $\mathbf{p}_{0}$ and $\mathbf{p}_{1}$ are vectors with the $x$ and $y$ coordinates of the two corner points of the wall segment, respectively.
 \item If necessary, define additional, site-specific scattering areas that are not aligned with a wall. This could be areas with large signs or other scattering objects of significance.
 \item Discard all scatterers drawn in step 1 that are not located within a scattering area.  
 \end{enumerate}
This simple algorithm is essentially a rejection sampler that ensures that the scatterers are placed uniformly within each scattering area.
The narrow and T-shaped intersection only contains the wall-type scatterers. The wide intersection has an additional diffuse scattering band in the middle of the widest road, motivated by the parked cars that are located there. The open intersection also has such bands in the middle of one road, containing non-wall scatterers such as large signs and lamp posts. The open intersection also includes areas that are obstructed by foliage and other objects, as indicated by dashed red lines. These areas should ideally be filled with diffuse scatterers, but it turns out that the contribution from them are so small that they can be neglected in this particular case.

\subsection{Path gain model}\label{sec:pathgain}
The path gain model for the different MPCs needs to include the effects of i) distance dependence, ii) losses due to interactions with the scattering objects, iii) obstructions by buildings, foliage and other objects, iv) diffraction around corners, v) angular dependence of the scattering interaction, vi) random large scale fading. We accomplish this by modelling the path gain with a classical log-distance power law to account for the distance dependence, and introduce additional factors for the remaining effects. In linear scale, the average path power gain, $\bar{g}^2$ for the each MPC is modelled as

\begin{align}
\bar{g}(d)^2=\left(\frac{g_{0}g_{a}g_{b}}{d}\right)^210^{-\frac{L_{\mathrm{p}}}{10}}.
\end{align}
Here, $d$ is the path propagation distance and $g_{0}^2$ is the path power gain at a reference distance of 1 m, assuming that $g_{a}=g_{b}=1$ and $L_{p}=0$. For the line-of-sight (LOS) component, $g_{0}^2$ is given by the free space path loss at a distance of 1 m.  The term $g_{a}^2$ is the path angular power gain, which is a function of the incoming and outgoing angles at each scatterer. The model framework does support the use of measured or theoretical angular gain functions for different types of scatterers. If this is to be used, we note that it is necessary to only use the envelope of such functions, since $g_{a}$ is meant to capture the average path gain; any random variation is captured by the fading term described in Sec. \ref{sec:fading}. In this paper, the following empirical expression is used to enable a simple parameterization of the model: 

\begin{eqnarray}
g_{a}=\mathrm{e}^{-\xi(|\theta_{1}-\theta_2|-\Delta\theta_{1})I_{1}-\xi(|\theta_{1}|-\Delta\theta_{2})I_{2}-\xi(|\theta_{2}|-\Delta\theta_{2})I_{3}},\label{eq:ga}
\end{eqnarray} 
where
\begin{eqnarray}
I_{1}&=& 
\begin{cases}
    0,& \text{if } |\theta_{1}-\theta_2|>\Delta\theta_{1}\\
    1,              & \text{otherwise}
\end{cases} \\
I_{2}&=& 
\begin{cases}
    0,& \text{if } |\theta_{1}|>\Delta\theta_{2}\\
    1,              & \text{otherwise}
\end{cases} \\
I_{3}&=& 
\begin{cases}
    0,& \text{if } |\theta_{2}|>\Delta\theta_{2}\\
    1,              & \text{otherwise}.
\end{cases} 
\end{eqnarray}
Here, $\theta_{1}$ and $\theta_{2}$ are the incoming and outgoing angles with respect the unit surface normal, $\hat{n}$, which is assigned to each scatter, and $\xi$ is an angular decay factor. $\Delta\theta_{1}$ and $\Delta\theta_{2}$ are constants that determine the angle region in which a path is unaffected by the angular decay $\xi$. We note that, based on our measurement results, we are only able to provide a very rough estimate for the parameter $\Delta\theta_{2}$. For scatterers associated with flat walls and signs, $\hat{n}$ is a usually unit normal vector pointing outward from the flat surface. Other, less well defined scatterers and diffuse scatterers can be assigned random normal unit vectors. The definition of $\theta_{1}$ and $\theta_{2}$ is given by Fig.~\ref{fig:angle}. 

\begin{figure}[h]
\centering
\begin{tikzpicture}

    \coordinate (O) at (0,0) ;
    \coordinate (A) at (0,2) ;
    \coordinate (B) at (0,-1) ;
    
    \fill[blue!25!,opacity=.3] (-2,0) rectangle (2,2);
    \fill[darkgray!60!,opacity=.3] (-2,0) rectangle (2,-1);
    \node[right] at (2,1) {};
    \node[left] at (0,-0.5) {Wall};

    \draw[dash pattern=on5pt off3pt] (A) -- (B) ;

    \draw[blue,ultra thick,reverse directed] (O) -- (130:2.6);
    \draw[blue,directed,ultra thick] (O) -- (45:2.7);
    \draw[black,normal,ultra thick] (O) -- (90:1.7);

    \draw (0,1) arc (90:130:1);
    \draw (0,1) arc (90:45:1) ;
    \node[] at (67:1.3)  {$\theta_{2}$};
    \node[] at (110:1.3)  {$\theta_{1}$};
    \node[] at (97:1.8)  {$\hat{n}$};
\end{tikzpicture}
\caption{Definition of incoming and outgoing angles at a scattering object. In this case, the scatterer is a wall, but this definition applies to all scatterers.}
\label{fig:angle}
\end{figure}
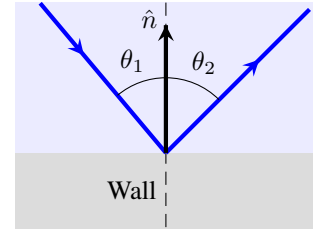

Let $\mathbf{a}=[a_{x}, \ a_{y}]$ be a vector pointing in the direction of the outgoing path in Fig.~\ref{fig:angle}, and let $\mathbf{b}=[b_{x}, \ b_{y}]$ be a vector pointing in the direction of the incoming path. In order to retain the proper sign of each angle, the angles can then be calculated as
\begin{eqnarray}
\theta_{1}=\mathrm{arctan}\left(\frac{b_{y}n_{x}-b_{x}n_{y}}{b_{x}n_{x}+b_{y}n_{y}}\right), \\
\theta_{2}=\mathrm{arctan}\left(\frac{a_{x}n_{y}-a_{y}n_{x}}{a_{x}n_{x}+a_{y}n_{y}}\right).
\end{eqnarray}
When using these equations, it is important to use the four quadrant version of the arctangent function to retain the correct angle values, and also to use these angles in radians, when calculating $g_{a}(\theta_{1},\theta_{2})$. An example of this path voltage gain function is shown in Fig.~\ref{fig:gain}.     
\begin{figure}[h]
	\centering
	\includegraphics[trim=0 0 50 0,clip,width=0.39\textwidth]{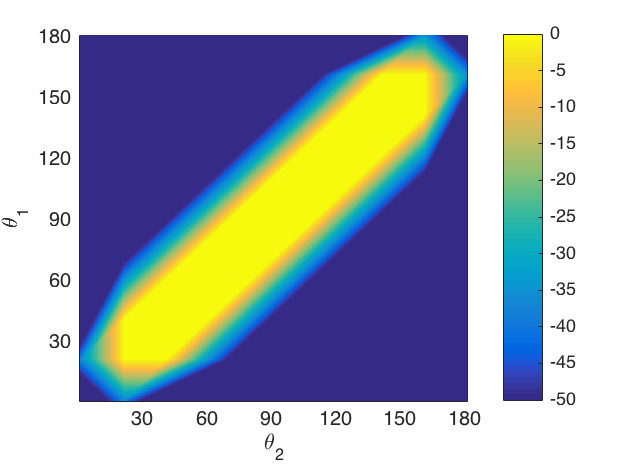}
	\caption{Angular path voltage gain function, $g_{a}(\theta_{1},\theta_{2})$, with parameters $\xi=12$ rad$^{-1}$, $\Delta\theta_{1}=0.35$ rad, and $\Delta\theta_{2}=1.22$ rad.}
	\label{fig:gain}
\end{figure}

Next, $g_{b}$ is a gain describing the effects of obstruction and blockage by buildings. Except for the LOS component, this term is simply an indicator function: $g_{b}=1$ if the path is not obstructed by any building, and $g_{b}=0$ if the path is obstructed. This choice is motivated by our measurements, which show that components that are obstructed by buildings disappear very rapidly, and that it makes the implementation less complex. For the LOS component on the other hand, it is necessary to include the effects of diffraction. A simple knife-edge diffraction model \cite{ITU} is applied, where the diffraction loss in dB is given by 

\begin{eqnarray*}
L_{d}(\nu)=6.9 + 20 \mathrm{log}_{10}\left(\sqrt{(\nu-0.1)^2+1}+\nu-0.1\right).
\end{eqnarray*}
The above applies if $\nu>-0.7$, otherwise, $L_{d}=0$ dB. Here, the term $\nu$ is given by
\begin{eqnarray}
\nu=\phi\sqrt{\frac{2}{\lambda\left(\frac{1}{d_{1}}+\frac{1}{d_{2}}\right)}},
\end{eqnarray}
where $\phi$ is the angle of diffraction, $d_{1}$ is the distance from the Tx to the building corner and $d_{2}$ is the distance from the corner to the Rx. So for the LOS component, $g_{b}=10^{-\frac{L_{d}}{20}}$.


Lastly, the model also supports additional losses due to areas with dense foliage, or areas with other objects that obstruct the path but do not completely block it. The measured data does not support the estimation of this blockage, so we instead opt to use an existing model for blockage due to foliage \cite{Marcus}. The penetration loss in dB is given by
\begin{eqnarray}
L_{p} = 0.2\left(f\cdot10^{-6}\right)^{0.3}d_{p}^{0.6},
\end{eqnarray}
where $d_{p}$ is the distance travelled through the obstruction area. We note that not all areas with foliage cause losses. For instance, alleys with trees might not be a significant issue if the tree canopies are situated significantly higher compared to the height of the car antennas. In the four intersections investigated in this paper, only the open intersection contains areas with additional losses. 

\subsection{Path gain parameter estimation}
Using the measured time-power-delay trajectories and the estimated location of each scatterer, we can now estimate the parameters of our average path power model for all of the measured scatterers. We use a maximum-likelihood estimator to jointly estimate $G_{0}=20\mathrm{log}_{10}(g_{0})$, $\xi$ and $\Delta\theta_{1}$. $G_{0}$ is found to be in the range of about $-48$ to $-75$ dB, with weaker powers for higher order interactions. The angular decay $\xi$ is estimated to be in the range from about 3-24 rad$^{-1}$, and $\Delta\theta_{1}$ is about 0.14-0.54 rad. To make the model simpler, we have chosen to use fixed values for these parameters, with $\xi=12$ rad$^{-1}$, $\Delta\theta_{1}=0.35$ rad and $\Delta\theta_{2}=1.22$ rad. A summary of all model parameters are found in Table II and III. Fig.~\ref{fig:gain2} shows an example of the estimated average path power gain of a first order MPC as a function of propagation distance. The power also depends on $\theta_{1}$ and $\theta_{2}$ through $g_{a}$. The path is characterized by a visible region with no building obstructions, and an obstructed region. In the obstructed region, only noise was being measured. In the visible region, there is an active region, which occurs when $I_1=I_2=I_3=0$, as given by (6)-(8). In this active region, the path decays with a slope of 2. Outside the active region, the path is affected by an additional decay due to the factor $g_{a}$, with an angular decay factor, $\xi$. For this specific path, the estimated decay is approximately $ \hat{\xi}=3$  rad$^{-1}$ on one side and $\hat{\xi}$ = 12 rad$^{-1}$ on the other side. However, as most paths have estimated angular decays that are very similar, the value is fixed to $\xi=12$ rad$^{-1}$ in the model.
\begin{figure}[h]
	\centering
	\includegraphics[width=0.48\textwidth]{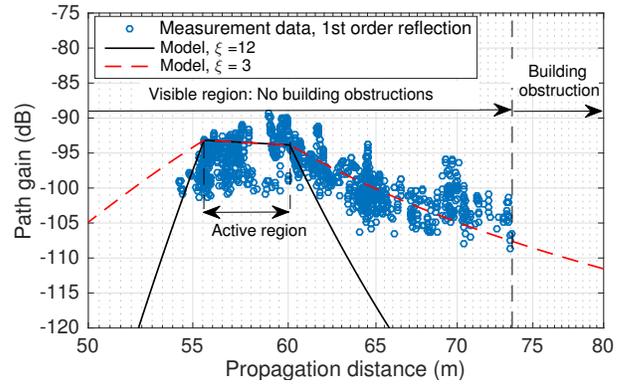}
	\caption{Estimated average path power gain of a first order MPC.}
	\label{fig:gain2}
\end{figure}

\subsection{Fading}\label{sec:fading}
The instantaneous path power gain with random shadow fading is then given by 
\begin{eqnarray}
g_{l}^2=\bar{g}(d_{l})^2\Psi_{\sigma_{\mathrm{s}}},
\end{eqnarray}
where $\Psi_{\sigma_{\mathrm{s}}}$ is a random variable that describes the fading about the distance-dependent mean path gain. This random fading is modelled as being Gamma-distributed, with parameters $k$ and $\theta$, and probability density function (PDF)
\begin{eqnarray}
f(x;k,\theta) = \frac{1}{\Gamma(k)\theta^k}x^{k-1}\mathrm{e}^{-x/\theta}, \ x,k,\theta > 0.
\end{eqnarray}

The average path gain power is given by $\bar{g}^2$, so the average power of $\Psi$ is always unity-mean, meaning that $\theta=1/k$, such that $\mathbb{E}[\Psi]=k\theta=1$. The parameters are estimated by a maximum-likelihood estimator, and $k$ is found to be in the range of $0.9-7.8$. These values are approximately uniformly distributed for the different scatterers. Model parameter values are given in Table II and III. 
\begin{figure}[h]
	\centering
	\includegraphics[width=0.48\textwidth]{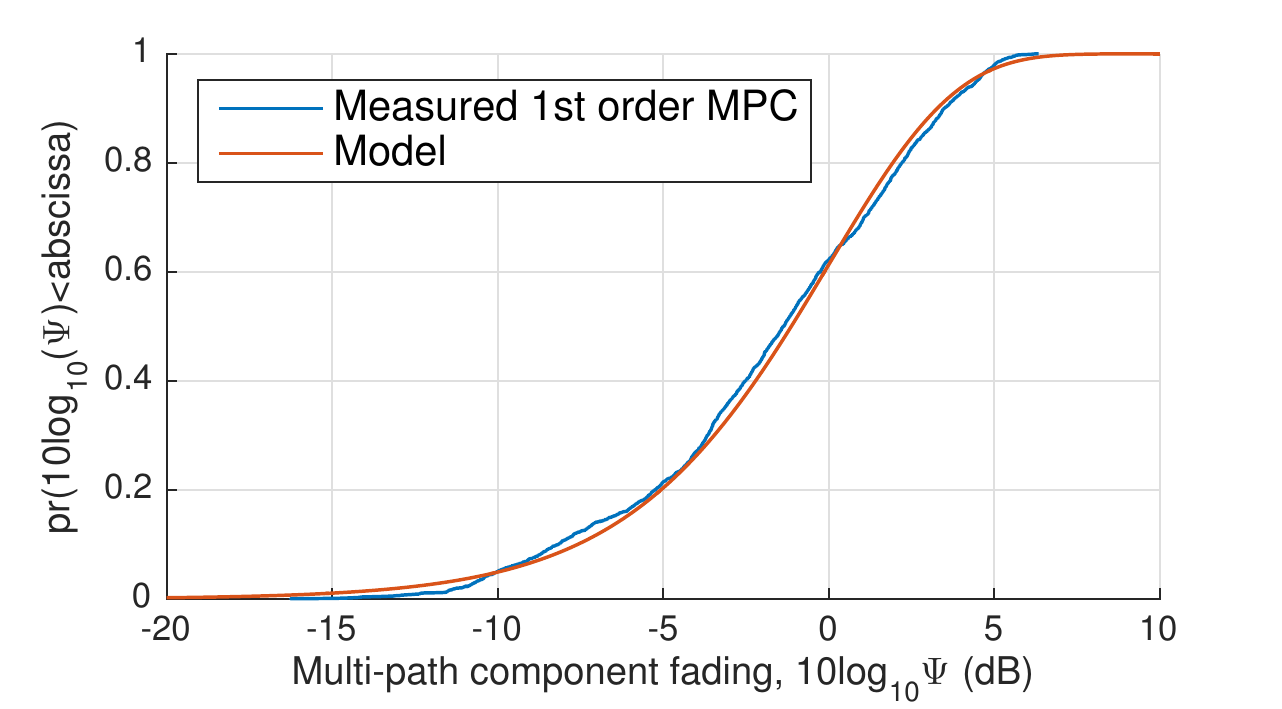}
	\caption{Empirical CDF of the fading, $\Psi$, of a measured first order MPC with parameters $k=1.36$ and $\theta=0.73$.}
	\label{fig:fading}
\end{figure}
As the power is Gamma-distributed, it also follows that the amplitude is Nakagami-distributed. Its PDF is given by
\begin{eqnarray}
f(x;m,\Omega) = \frac{2m^m}{\Gamma(m)\Omega^m}x^{2m-1}\mathrm{e}^{-\frac{m}{\Omega}x^2},
\end{eqnarray}
where $m=k$ and $\Omega=m\theta$. The Nakagami distribution with $m=1$ is identical to the Rayleigh distribution, and for $m>1$, the Nakagami distribution can approximate the Rice distribution\cite{Braun}, although with a different slope close to $x = 0$, which impacts the achievable diversity order \cite{Molisch}. One physical interpretation of this is that the amplitude fading of each MPC is caused by a vector process consisting of several scattering contributions with similar delays. Nakagami fading can be employed for cases when the central limit theorem is not necessarily valid, which can be the case for ultra-wideband channels \cite{Molisch}. A benefit of the Nakagami distribution is that both Rayleigh and Rice-like fading can be emulated with a single distribution. For instance, an MPC with $m \approx 1$  is made up of a number of components of similar strength, whereas an MPC with $m>1$ also contains a dominating component. Our estimates of $m$ indicates that higher order interactions generally have smaller values of $m$ compared to that of first-order interactions. This seems reasonable, as paths with higher interaction orders are by nature less likely to contain one dominating component. Fig.~\ref{fig:fading} shows cumulative distribution functions of the measured and estimated fading for a measured first order MPC.  
\subsection{Small-scale fading due to multi-path interference}
As the received signal is composed of a summation of several MPCs, the resulting fading in each delay bin will also depend on the bandwidth used by the communication system. The GSCM is capable of modelling this dependence on bandwidth, as the fading in each delay bin is modelled as a summation of MPCs with different Nakagami-distributed amplitudes and different random phases. The fading in each delay bin will therefore be Rice- or Rayleigh-like, just as reported in \cite{Karedal}.

\subsection{Autocorrelation}
The random fading term, $\Psi_{\sigma}$, is an autocorrelated Gamma-fading process, which we describe based on the classical Gudmundson model, i.e., the auto-correlation function is modelled as 

\begin{eqnarray}
r(\Delta d)_{l}=\sigma^2_{l}\mathrm{e}^{-|\Delta d|/d_{c,l}},
\end{eqnarray}
where $\sigma^2_{l}$ is the variance of the fading process and $d_{c,l}$ is the coherence distance. Fig.~\ref{fig:auto1} shows sample autocorrelation functions for five different paths. Fig.~\ref{fig:auto2} shows the same thing for a single path, but also shows the estimated  exponential autocorrelation function with an estimated coherence distance of 0.9 m. The estimated coherence distance for different paths is found to be in the range of about 0-1.9 m.

\begin{figure}[h]
	\centering
	\includegraphics[width=0.45\textwidth]{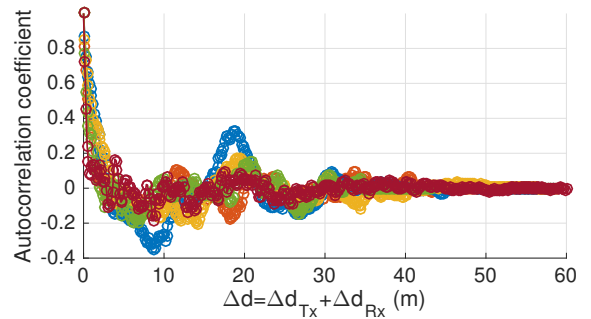}
	\caption{Sample autocorrelation functions for five paths.}
	\label{fig:auto1}
\end{figure}

\begin{figure}[h]
	\centering
	\includegraphics[width=0.45\textwidth]{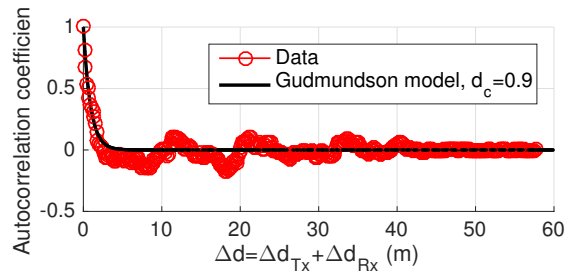}
	\caption{Sample autocorrelation coefficient for a single path, and the modelled autocorrelation with an estimated coherence distance of 0.9~m.}
	\label{fig:auto2}
\end{figure}

This autocorrelated Gamma-process can be implemented in various ways. We use the following approximate method based on a numerical discretization of an It\^o form stochastic differential equation \cite{Dima}. Sample number $u+1$ for the realization $\Psi_{u+1}$ is generated by:
\begin{eqnarray}
\frac{\Psi_{u}d_{c}+u\theta\Delta d+\theta\Delta d(\xi_{u}^2-1)/2+\sqrt{2\Psi_{u}\theta d_c\Delta d}\xi_{u}}{\Delta d + d_c}. \label{Eq.Ito}
\end{eqnarray}
The process is then generated as follows: 
\begin{enumerate}
\item Generate an initial value, $\Psi_0 \sim$ Gamma$(k,\theta)$. 
\item Calculate the total distances moved by the two cars during the time between neighbouring samples $\Psi_{u}$ and $\Psi_{u+1}$:  $\Delta d_{u,u+1} = \Delta d_{\mathrm{Tx},u,u+1} + \Delta d_{\mathrm{Rx},u,u+1}$.
\item Generate $\xi_{u}$ drawn from the standard normal distribution.
\item Generate $\Psi_{u}$ according to (\ref{Eq.Ito}) for samples $u=1,2,\ldots, U-1$, where $U$ it the desired number of samples.
\end{enumerate}
 
\subsection{Generating Channel Matrices}
We can now model the time-variant, double-directional and complex channel frequency transfer function as a superposition of $L$ different multi-path components \cite{DD,Karedal}:
\begin{eqnarray}
H(f,t)= \sum_{l=1}^Lg_{l}\mathrm{e}^{-\mathrm{j}2\pi f\tau_{l}}G_{\mathrm{Tx}}(\Omega_{\mathrm{Tx}})G_{\mathrm{Rx}}(\Omega_{\mathrm{Rx}}). \label{Eq.H}
\end{eqnarray}
Here, $G$ is the complex antenna amplitude gain in direction $\Omega$, $g_{l}$ is the complex amplitude of path $l$ given by (1), $L$ is the total number of paths and $f$ is the frequency. Lastly, $\tau_l$ is the propagation delay, given by $\tau_{l} = d_{l}/c$. It is then possible to use (\ref{Eq.H}) to emulate MIMO channel matrices by simply summing up the received paths at each antenna element, while taking into account the path distance between each antenna pair. This way, the difference in small scale fading experienced by each antenna pair is being modelled by the constructive and destructive interference of all the paths, without having to derive any specific distribution for the small scale fading. We note that if multiple antennas are placed far away from each other on the car body, they will not experience the same large-scale fading realization. As the path coherence distance has been found to be on the order of 0-2 m, the path gain for distributed antenna elements can in fact experience different realizations of $g_{l}$ at the same time instant \cite{AbbasAntenna}.    

\begin{figure*}[t]
    \centering
        \includegraphics[height=3.5cm]{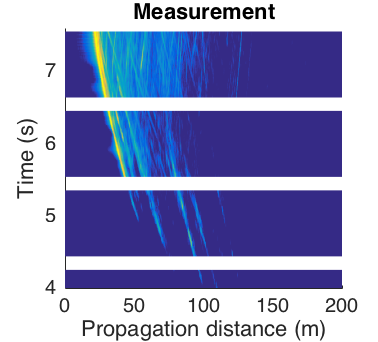}%
        \hfill
        \includegraphics[height=3.5cm]{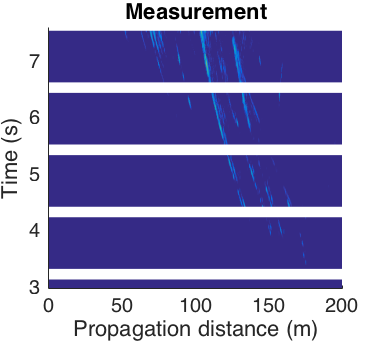}
        \hfill
        \includegraphics[height=3.5cm]{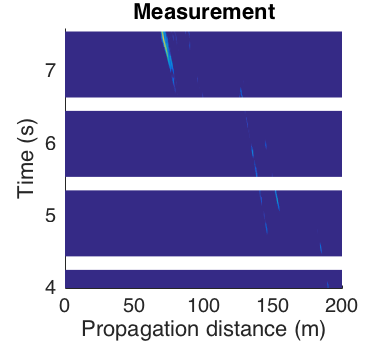}
        \hfill
        \includegraphics[height=3.5cm]{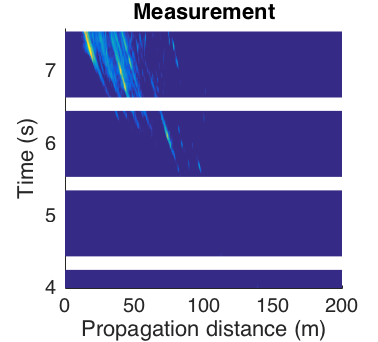}
        \hfill
        \includegraphics[trim=320 0 0 0,clip,height=3.5cm]{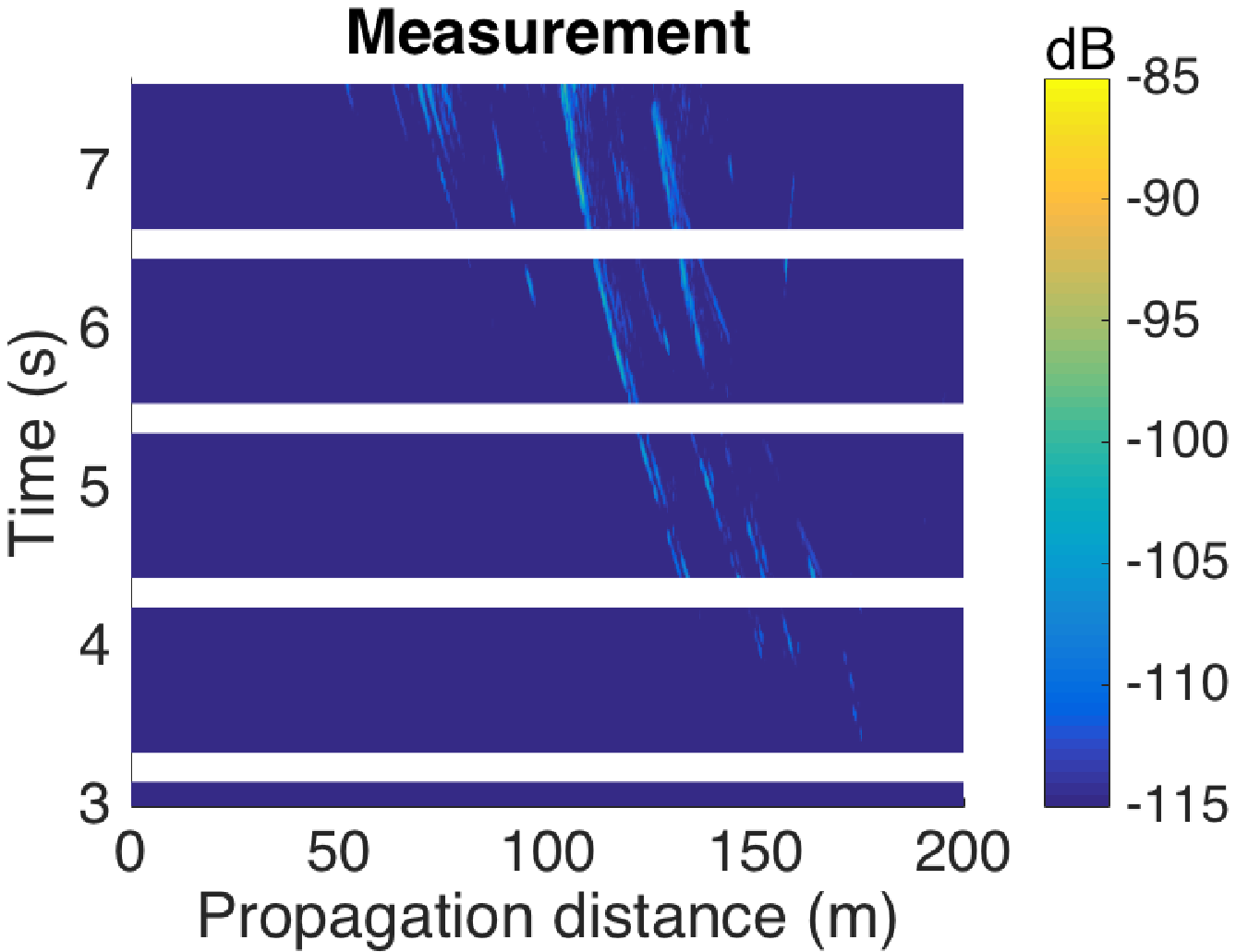} \\
        \includegraphics[height=3.5cm]{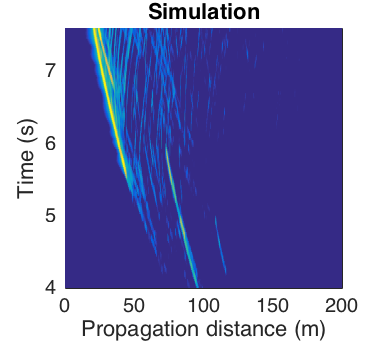}%
        \hfill
        \includegraphics[height=3.5cm]{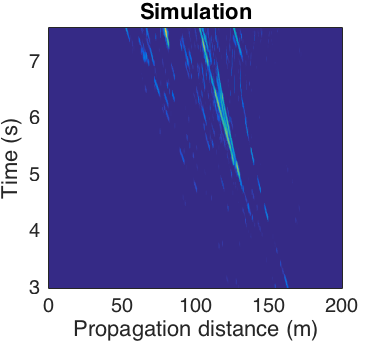}
        \hfill
        \includegraphics[height=3.5cm]{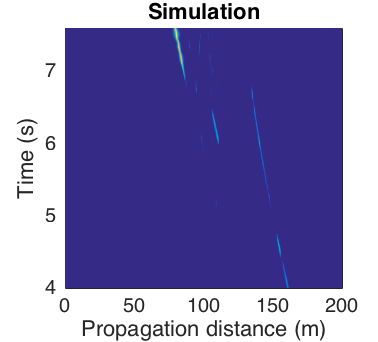}
                \hfill
        \includegraphics[height=3.5cm]{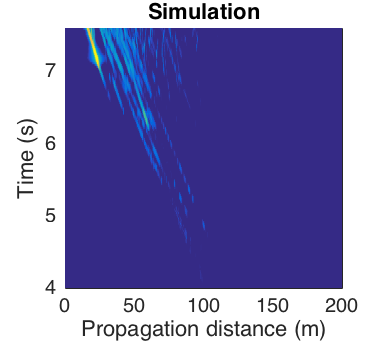}
        \hfill
        \includegraphics[trim=320 0 0 0,clip,height=3.5cm]{Colorbar.eps} 
    \label{fig:PDP}
    \caption{Measured (top row) and simulated (bottom row) power-delay profiles, with power in dB, as a function of time and path propagation distance, for the narrow, wide, open, and T-shaped intersections. This illustrates the capabilities of the GSCM model in terms of capturing the general behavior of the propagation channel in various types of intersections. The gaps in the time axes for the measurements are due to a limitation of the channel sounder used in the measurement.}
\end{figure*}

Similar to the approach in \cite{Karedal}, we model (\ref{Eq.H}) using six different parts; the line-of-sight path (LOS), first, second and third order wall reflections, first order reflections from non-wall objects and finally first order constributions from diffuse interactions. It is easy to add mobile scatterers as well, but this has been omitted, as the measurements indicated that such components, with a significant strength, seldom appear in the measured scenarios. The total transfer function for a single Tx-Rx antenna pair can then be calculated using Eq.~ \ref{Eq.H} as: 

\begin{eqnarray}
H(f,t)_{\mathrm{tot}}= g_{\mathrm{LOS}}\mathrm{e}^{-\mathrm{j}2\pi f\tau_{\mathrm{LOS}}} + \sum_{w_{1}=1}^{W_1}g_{w_{1}}\mathrm{e}^{-\mathrm{j}2\pi f\tau_{w_{1}}} \nonumber \\
+\sum_{w_{2}=1}^{W_2}g_{w_{2}}\mathrm{e}^{-\mathrm{j}2\pi f\tau_{w_{2}}}
+\sum_{w_{3}=1}^{W_3}g_{w_{3}}\mathrm{e}^{-\mathrm{j}2\pi f\tau_{w_{3}}} \nonumber \\
+\sum_{s=1}^{S}g_{s}\mathrm{e}^{-\mathrm{j}2\pi f\tau_{s}}
+\sum_{di=1}^{Di}g_{di}\mathrm{e}^{-\mathrm{j}2\pi f\tau_{di}}. \label{Eq.Htot}
\end{eqnarray}
 Here, the influence of the antenna patterns in (\ref{Eq.H}) has been omitted for clarity. A MIMO channel matrix $\mathbf{H}(f,t)$ can be generated by using (19) while taking into account the difference in delay, $\tau$, for the different antenna combinations.



\subsection{Model parameters}
Table II and III summarize the most important parameters for the GSCM. These are the parameters that have been used in our simulations. The path gain is specified for first, second and third order interactions for scatterers placed along walls as well as diffuse and non-wall first order scatterers. The diffuse scatterers are also placed along walls, in bands with a wider width of 12 m (unless the width of half the street is less than 12 m, in which case the scatterers are placed all over the whole street). Diffuse and non-wall scatterers might also be placed in user defined polygons for scattering areas that are not aligned with any walls. This mostly applies for wider and more open intersections.   
\begin{table}[htp]
\caption{Path gain parameters}
\begin{center}
\begin{tabular}{|c|c|c|c|c|c|} \hline
Type & Order &$G_{0}$ (dB)  & $d_{c}$ (m) & $k$ & $\theta$ \\ \hline
Wall &1st&$U(-65,-48)$ & $U(1,2)$& $U(2,8)$ & $1/k$  \\ \hline
Wall &2nd&$U(-70,-59)$ & $U(0,1.5)$& $U(1,6)$ & $1/k$ \\ \hline
Wall &3rd&$U(-75,-65)$ & $U(0,1)$& $U(1,4)$ & $1/k$ \\ \hline
Non-wall&1st&$U(-68,-52)$ & $U(0,1)$& $U(1,6)$ & $1/k$ \\ \hline
Diffuse&1st&$U(-80,-68)$ & $U(0,1)$& $U(1,1)$ & $1/k$ \\ \hline \hline
Type & Order & $\Delta\theta_{1}$ (rad) & $\Delta\theta_{2}$ (rad) & $\xi$ (rad$^{-1}$) & \\ \hline
All & All & 0.35 & 1.22 & 12 & \\ \hline
\end{tabular}
\end{center}
\label{default}
\end{table}%

\begin{table}[htp]
\caption{Scatterer location parameters}
\begin{center}
\begin{tabular}{|c|c|c|c|c|c|c|} \hline
Type &Order& $\chi$ (m$^{-2}$) & $W$ (m) \\ \hline
Wall&1st& 0.044 & 3 \\ \hline
Wall&2nd& 0.044 & 3 \\ \hline
Wall&3rd& 0.044 & 3 \\ \hline
Diffuse, wall&1st& 0.61 & 12  \\ \hline
Non-wall&1st& 0.034 & User defined \\ \hline
Diffuse, non-wall&1st& 0.61 & User defined \\ \hline
\end{tabular}
\end{center}
\label{default}
\end{table}%

\section{Results and Validation}
The model needs to be validated against measurement data to ensure that it gives reasonable results. The model parameters have been estimated for the most part on what we refer to as the narrow intersection, and for some parts based on the open intersection. Using the parameters given in Table II and III, we now simulate channel matrices using the developed GSCM model, for the narrow, wide, open and T-shaped intersections, based on parameters extracted from the narrow and open intersections. 
For the validation, an inverse Fourier transform is performed over the frequency domain of the simulated channel transfer functions $H(f,t)$, to obtain the impulse responses $h(\tau,t)$. These responses are then used to derive power-delay profiles (PDPs) by averaging over a window of $N$ time instants:

\begin{eqnarray}
P(\tau,t_{a}) = \frac{1}{N}\sum^{N-1}_{n=0}|h(t_{a}+n\Delta t,\tau)|^2.
\end{eqnarray}
Here, $\Delta t$ is the time difference between two consecutive data samples and $N$ is chosen such that $N\Delta t= 30$ ms. At a speed of 50 km/h, this corresponds to a movement of about 8 wavelengths, or 0.42 m. This distance is typically within the estimated coherence distance of most of the significant paths the we have observed in our measurements. It should be noted that there are paths that have a coherence distance which is shorter than our chosen stationary interval. However, these paths are very weak, and do not contribute significantly to the condensed channel parameters such as the power-delay profile, RMS delay spread and channel gain. The measured and simulated power-delay profiles (PDPs) are shown in Fig.~11, illustrating that the developed GSCM is capable of reproducing the overall propagation behavior in these four different intersection. The most notable difference is the somewhat larger portion of dense diffuse scatterers present in the measured PDP in the narrow intersection, as compared to the simulated one.

To better quantify the propagation channel characteristics of the measurements and the simulations, we also present condensed channel parameters in terms of the channel gain, the mean delay and the RMS delay spread. The RMS delay spread is calculated as
\begin{eqnarray}
S(t_{a}) = \sqrt{\frac{\sum_{i}P(\tau_{i},t_{a})\tau_{i}^2}{\sum_{i}P(\tau_{i},t_{a})}-\left(\frac{\sum_{i}P(\tau_{i},t_{a})\tau_{i}}{\sum_{i}P(\tau_{i},t_{a})}\right)^2}.
\end{eqnarray}

The channel gain is calculated by summing up the noise-free power contributions in the PDP (any contribution with power less than 5 dB above the noise floor is discarded), 
\begin{eqnarray}
G(t_{a}) = \sum_{\tau}P(\tau,t_{a}).
\end{eqnarray}

\begin{figure}[h!]
	\centering
	\includegraphics[width=0.23\textwidth]{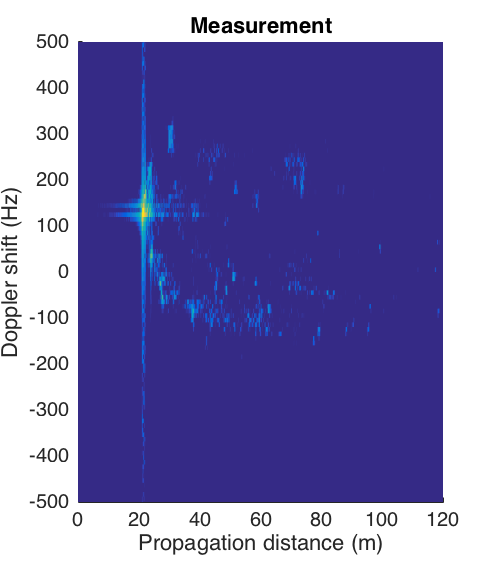}
	\includegraphics[width=0.23\textwidth]{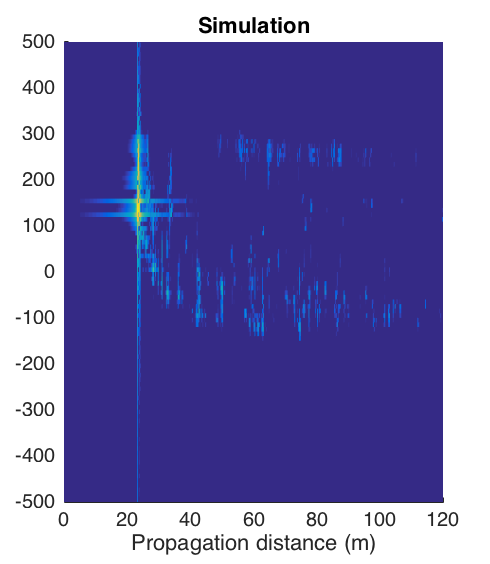}
	\caption{Example of a measured and a simulated Doppler-resolved impulse response. This is derived using a time window of 0.1 s, at 7.0-7.1 s of the measurement in the narrow intersection shown in Fig.~11.}
	\label{fig:dopp}
\end{figure}

Fig.~\ref{fig:dopp} shows the Doppler-resolved impulse response, $h(\nu,\tau)$, from a measurement and from a simulation, illustrating that the overall Doppler behavior is captured by the model. The Doppler response is derived by Fourier transforming the impulse responses with respect to time. In our example, this is done for the narrow intersection, using a time window of 7.0-7.1 s.  Looking at the details, it is clear that the measured Doppler shifts are slightly more concentrated compared to the simulation. This is likely not an issue for most analyses, since the resulting Doppler spreads are comparable.

\begin{figure*}[t]
        \centering
        \includegraphics[height=4.4cm]{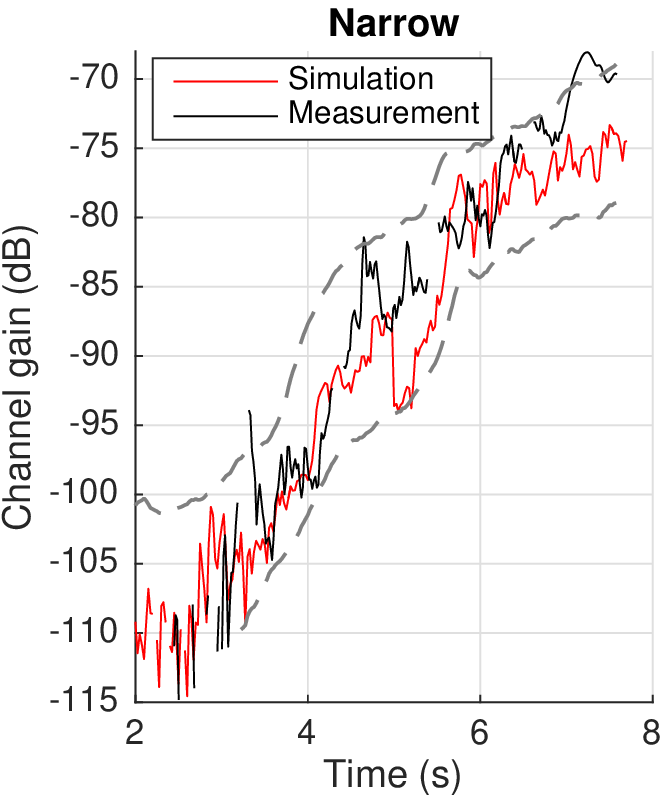}%
        \hfill
        \includegraphics[height=4.4cm]{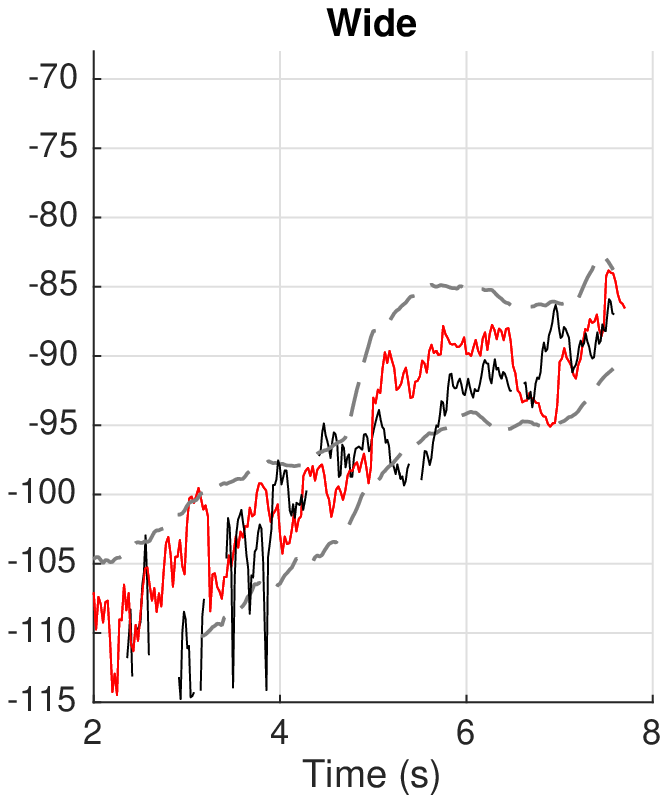}
        \hfill
        \includegraphics[height=4.4cm]{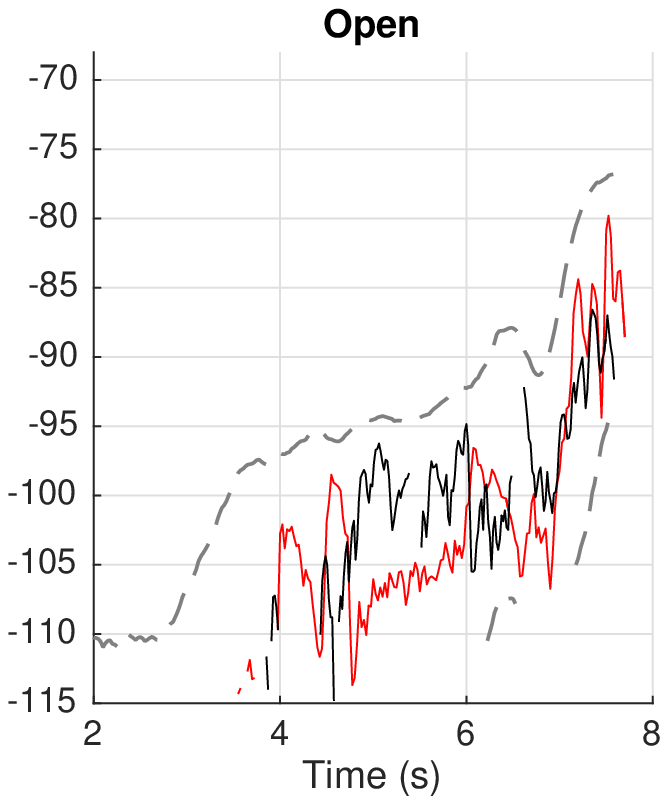}
        \hfill
        \includegraphics[height=4.4cm]{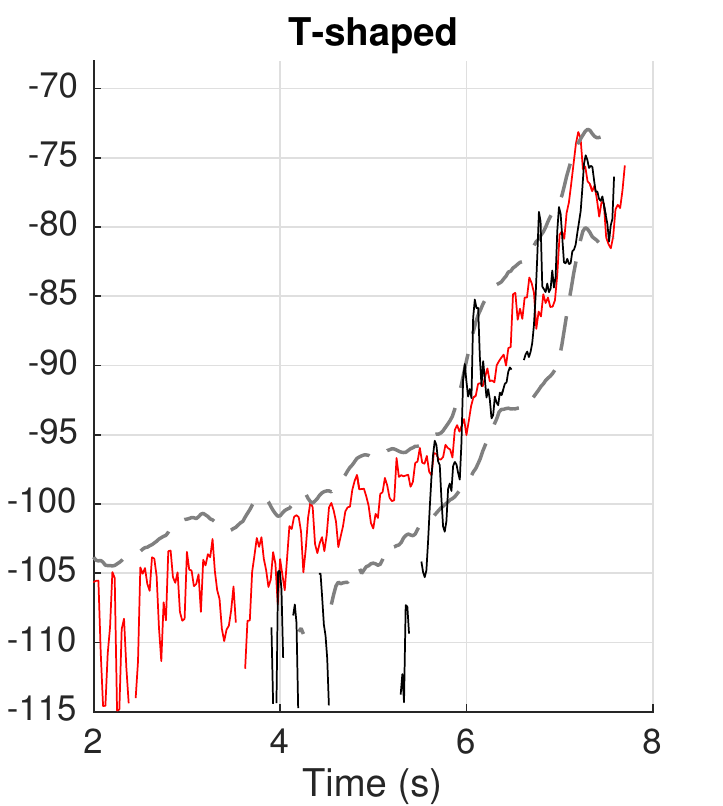}
        \includegraphics[height=4.4cm]{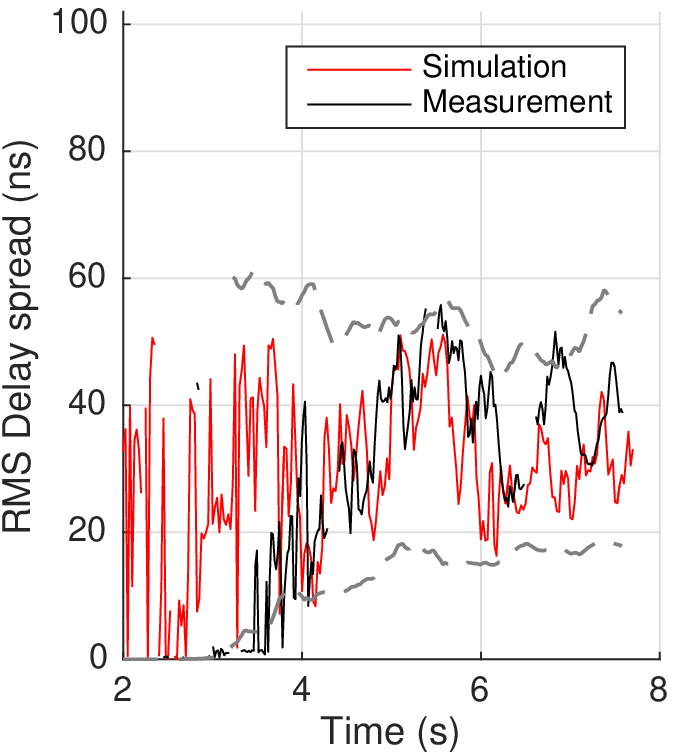}%
        \hfill
        \includegraphics[height=4.4cm]{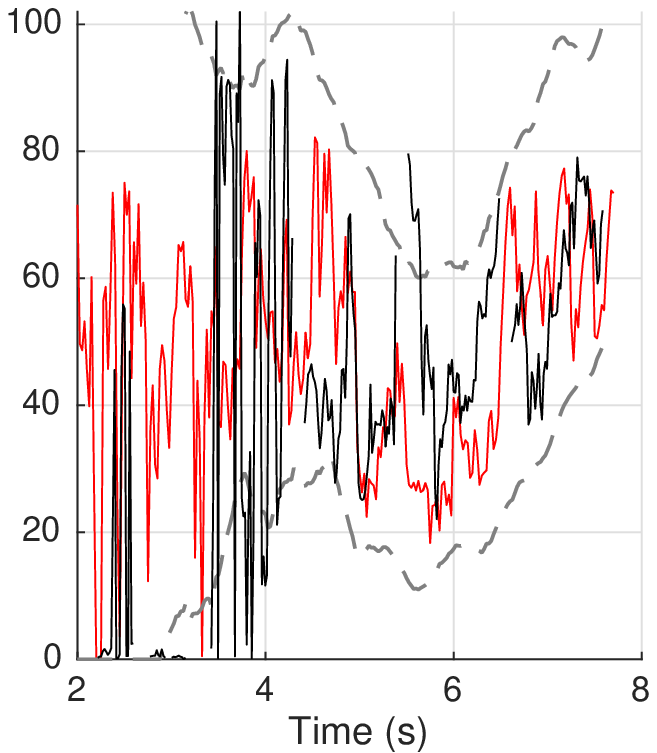}
        \hfill
        \includegraphics[height=4.4cm]{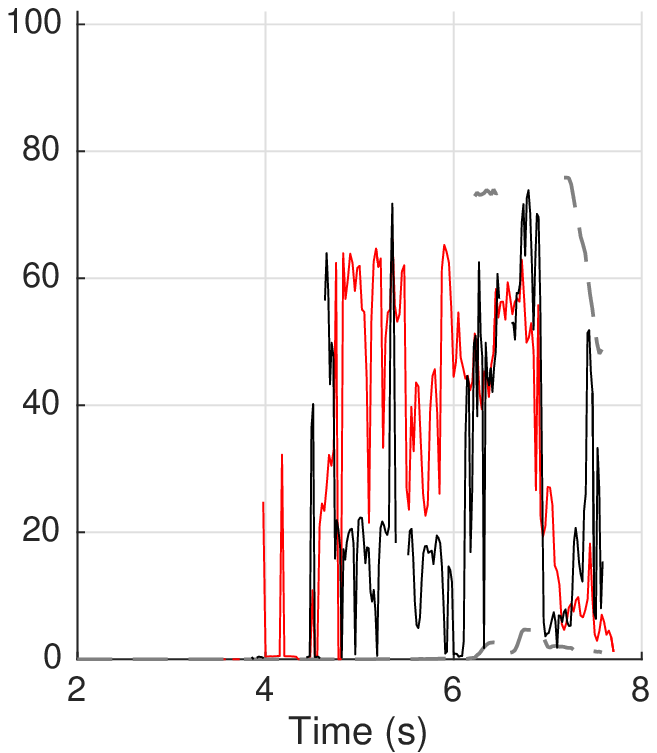}
        \hfill
        \includegraphics[height=4.4cm]{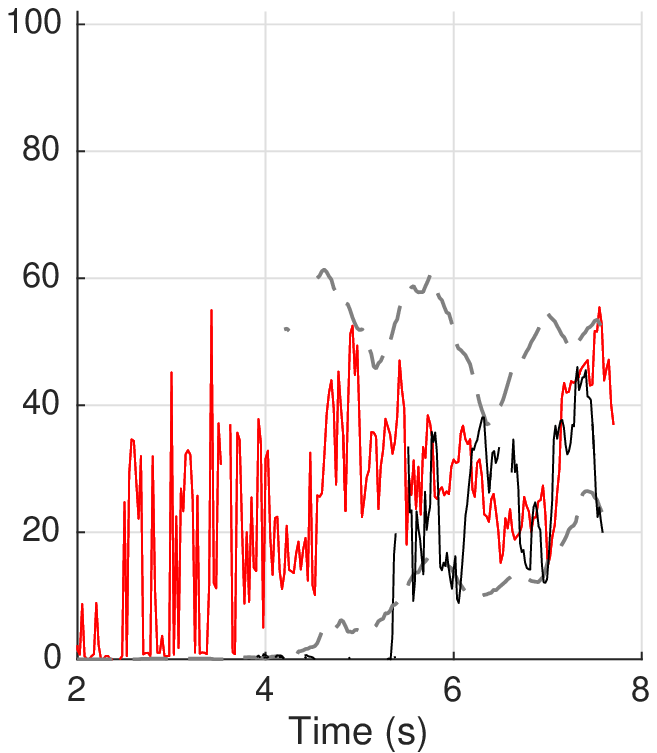}
\label{fig:simmeas}
    \caption{The GSCM is able to accurately capture the general behavior of both the channel gain and the RMS delay spread as a function of time in all four intersections. The dashed gray lines represent approximate bounds for largest and smallest values taken from 100 separate realizations.}
\end{figure*}

Lastly, Fig.~13 shows the measured and simulated channel gain, $G(t_{a})$, and RMS delay spread, $S(t_{a})$ for all four intersections. For the sake of clarity, we only present results for a single measurement run in each intersections, and for a single antenna combination. The results for the other measurement runs and antenna combinations are comparable to the ones shown in Fig.~13. The red lines are for a single simulation run. However, as the GSCM model is stochastic in nature, we also present approximate bounds for the smallest and largest values taken from 100 separate simulation runs, as indicated by the dashed gray lines. The simulated channel gain agrees very well with the measurements, and for the most part, the measurements lie within the expected bounds. A small discrepancy can be noticed at around 5 s for the T-shaped intersection. This is likely due to one or several objects blocking the signal pathways, that are not present in the simulation environment. 

It is also seen that the channel gain can vary drastically over time, but also across different types of intersections. The simulation of the RMS delay spread also show a good agreement with measurements. The wide intersection displays the largest delay spread, whereas the open intersections has a very small delay spread. This is because there are few significant components present in this specific measurement run in the open intersection, resulting in a small delay spread.

\section{Conclusion and Future work}
We have developed a novel GSCM capable of emulating the typical properties of V2V propagation channels in arbitrary urban environments. Model parameter estimates are given based on high-resolution measurements in two separate intersections in the city of Berlin, Germany; a narrow and an open-type intersection. Using these parameter estimates, the model is validated against measurement data from the narrow and open intersection, and against a wide and a T-shaped intersection. The results show that the model can accurately reflect the propagation channel properties in these four intersections without having to tune model parameters. This could indicate a certain model robustness, although further validation using measurement-based results from additional environments is still needed. For instance, an environment with very few scattering objects, and buildings with very flat facades might exhibit a different behavior.
However, the developed model is still able to accurately emulate many urban V2V scenarios. 

The most challenging types of environments are open-type intersections and intersections that have areas with vegetation, or other objects that might obstruct the MPCs. For these intersections, a lot of effort should be put in trying to model the behavior of the areas with obstructions. 
As future work, we intend to tune the model parameter estimates to additional measured urban environments, including environments with forestation or heavy foliage, and places with open water, such as wide canals. We also aim at extending the model framework to 3D scenarios, in order to support the modeling of vehicle-to-cellular channels and cellular channels above 6 GHz.

\section*{Acknowledgment}
This work has been developed within the framework of the COST Action CA15104, Inclusive Radio Communication Networks for 5G and Beyond (IRACON). Parts of the work have been funded by grants from FFI/Vinnova, and ELLIIT, the Excellence center at Link\"oping-Lund in Information Techn.

\ifCLASSOPTIONcaptionsoff
  \newpage
\fi

\bibliographystyle{IEEEtran}
\bibliography{Pathlossbib.bib}

\end{document}